\documentclass[12pt]{iopart}

\usepackage{setspace}
\usepackage{subcaption}
\usepackage{graphicx}
\usepackage{dcolumn}
\usepackage{bm}
\usepackage{hyperref}
\usepackage[dvipsnames]{xcolor}
\usepackage{url}
\usepackage[utf8]{inputenc}
\usepackage{tabularx}
\usepackage{tablefootnote}
\usepackage{cite}
\usepackage{lineno}
\begin{document}
\title[]{Jet modification in absence of QGP-medium: the role of multiparton interactions and color reconnection}

\author{Prottoy~Das$^{1,a}$, Abhi~Modak$^1$, Debjani~Banerjee$^1$, Rathijit~Biswas$^2$, Supriya~Das$^1$, Sanjay~K.~Ghosh$^1$, Sibaji~Raha$^1$, Sidharth~Kumar~Prasad$^{1,b}$}

\address{$^1$Department of Physics, Bose Institute, Kolkata - 700091 (INDIA)}
\address{$^2$Quark Matter Research Center, Institute of Modern Physics, Chinese Academy of Sciences, Lanzhou - 730000 (CHINA)}
\eads{\mailto{$^a$dasprottoy000@gmail.com},\mailto{$^b$sprasad@cern.ch}}
\vspace{10pt}
\begin{indented}
\item[]
\end{indented}

\begin{abstract}
  Recent studies of high-multiplicity events in small collision systems (proton-proton and proton-lead) have drawn research interest towards the possibility of the formation of partonic medium in such systems. One of the important consequences of the formation of dense partonic medium is quenching of high-momentum final-state particles resulting in several experimental observations such as suppression in nuclear modification factor $R_{\rm AA}$, modification of jet shape observable $\rho(r)$ and jet fragmentation ($z^{\rm ch}$) distributions, etc. In this work, we study $\rho(r)$ and $z^{\rm ch}$ for inclusive charged-particle jets in proton-proton (pp) collisions at $\sqrt{s}$~=~13~TeV using PYTHIA~8 Monash~2013 Monte Carlo simulation. We show that the color reconnection (CR) and multiparton interaction (MPI) mechanisms in PYTHIA~8 can lead to an increased rate of jet production. We also find that the mechanisms of MPI and CR and change in the gluonic contribution in high-multiplicity events result in significant modification of $\rho(r)$ and $z^{\rm ch}$ compared to those in minimum bias events for 10~$<p_{\rm T,\,jet}^{\rm ch}<$~20~GeV/$c$. We notice a direct connection of $\langle N_{\rm MPI}\rangle$ and gluonic contribution with the amount of modification in $\rho(r)$ -- the larger the number of MPIs and/or gluonic contribution, the larger the amount of modification of $\rho(r)$.
\end{abstract}


\section{\label{sec:Intro}Introduction}
Experimental observations at RHIC and LHC have revealed that a system of deconfined quarks and gluons, known as Quark-Gluon Plasma (QGP) is formed in heavy-ion (AA) collisions at relativistic speeds~\cite{ref-QGP-formation1STAR,ref-QGP-formation2STAR,ref-QGP-formation3STAR,ref-QGP-formation1ALICE,ref-QGP-formation1CMS,ref-QGP-formation2CMS,ref-QGP-formation1ATLAS}. A conclusive statement about the formation of QGP requires a comparison of measurements in heavy-ion collisions with those from proton-proton (pp) and proton-nucleus (pA) collisions at similar energies where conventionally no such system is expected to be present. Among others, jet quenching and collective effects are the two most striking observations whose presence in AA and absence in pp and pA collisions provide strong evidence of QGP formation~\cite{ref-jetQuenching-Flow-QGP}. In high-energy elementary, hadronic and nucleus-nucleus collisions, the processes with large (at a scale~$\textgreater \textgreater~\lambda_{\rm QCD}$) momentum transfer (large $Q^2$) result in two back-to-back high-momentum partons (quarks and gluons). These high-momentum partons lose their virtuality by producing a collimated cascade of partons in the direction of the parent parton. Eventually, all partons manifest themselves into collimated showers of experimentally detectable hadrons, known as jets, via soft hadronization processes. The jet production and its properties are well described by the theory of perturbative Quantum Chromodynamics (pQCD) in pp collisions~\cite{refJetPP,ALICEjetPP7TeV,ReviewJet}.
\par
Although theoretical calculations~\cite{ThQGPInSmallColl1,ThQGPInSmallColl2,ThQGPInSmallColl3,ThQGPInSmallColl4} hinted towards the possibility of collective effects in small collision systems long ago, with the onset of LHC, several experimental observations and model calculations could actually find signatures indicating the existence of these effects in pp and pA collisions, particularly at high multiplicities~\cite{small-sys-radial-flow1,small-sys-radial-flow2,small-sys-radial-flow3,small-sys-ridge1,small-sys-ridge2,small-sys-ridge3,small-sys-ridge4,small-sys-mass-ordering1,small-sys-mass-ordering2,small-sys-strangeness1,small-sys-strangeness2,CREffect1,CREffect2}. Interestingly no jet quenching has been reported in such events~\cite{small-sys-noJetQuenching1,small-sys-noJetQuenching2}. It has therefore drawn an immense interest towards the study and understanding of the dynamics of small collision systems on both theoretical and experimental fronts. Recently several efforts are made to understand the sources for the production of high-multiplicity events and their effects on the final-state particles~\cite{ref-pp-highmult1,ref-pp-highmult2,ref-pp-highmult3,ref-pp-highmult4,ref-pp-highmult5,ref-pp-highmult6,ref-pp-highmult7,ref-pp-highmult8,ref-pp-highmult9,ref-pp-highmult9v2,ref-pp-highmult10,ref-pp-highmult11,ref-pp-highmult12,ref-pp-highmult13,ref-pp-highmult14,ref-pp-highmult15,ref-pp-highmult16,resonance1,Ridge1-pp,Jpsi1,CGC3,CGC1,CGC2,CGC4,JetModInSmallSys1,JetModInSmallSys2,pajares1,pajares2,pajares3,pajares4,pajares5,pajares6,JetModGustafson,Zakharov1}. It is suggested in Refs.~\cite{CGC1,CGC2,CGC4,CGC5} that the existence of color-glass-condensate in the initial state can give rise to large particle production. The phenomena such as color reconnections (CR) between the outgoing partons and multiparton interactions (MPI) are also found to explain the production of large number of final-state particles in pp collisions~\cite{CREffect1,ref-pp-highmult9,ref-pp-highmult10,ref-pp-highmult11}. Ample signatures such as radial flow-like effects on spectra, ridge-like structure, mass ordering of elliptic flow ($v_{\rm 2}$), strangeness enhancement, etc., which are traditionally explained through the presence of hot and dense QGP medium in AA collisions, are also observed experimentally in high-multiplicity pp and pA collisions at LHC energies. Surprisingly, inclusive jet nuclear modification factor ($R_{\rm AA}$) is found to be unity thereby challenging the notion of medium formation in these systems. In AA collisions, among others, differential jet observables such as jet shape ($\rho(r)$) and jet fragmentation function ($z^{\rm ch}$) are studied to understand the jet-medium interaction in detail~\cite{CMS:Rho2pt76,CMS:Rho5pt02,ATLAS:JetFrag1,ATLAS:JetFrag2,ATLAS:JetFrag3,CMS:JetFrag1,CMS:JetFrag2}. Jets are found to get softer and broader in presence of medium due to elastic and inelastic collisions. Let us now concentrate on the present status of pp studies.
Interestingly, recent ALICE measurement~\cite{JetFragPreliDB} of jet fragmentation function ($z^{\rm ch}$) in pp collisions at $\sqrt{s}$ = 13 TeV shows significant modification at high multiplicity compared to minimum bias events. This is also reproduced by PYTHIA 8 Monash 2013 where no QGP medium effect is implemented. In Ref.~\cite{ref-pp-jetShape}, MPI and CR mechanisms of PYTHIA~8 are shown to modify jet shape ($\rho(r)$) distributions in high-multiplicity pp collisions, however, the exact interplay between these mechanisms causing the jet modification is not studied yet. Moreover, the modification of different jet observables may also have different degrees of sensitivity on the nature of parent parton. It is, therefore, of utmost importance to understand in detail and quantify various contributions from the individual sources leading to the jet modification in PYTHIA 8 in absence of the medium.

\par
In this work, we study the jet shape observable $\rho(r)$ and jet fragmentation function ($z^{\rm ch}$) in minimum bias and high-multiplicity event classes using PYTHIA~8 Monash~2013 MC event generator. The study is performed in presence and absence of the two important mechanisms in PYTHIA~8, CR and MPI (which are found to explain some of the experimentally observed flow-like features in high-multiplicity pp collisions~\cite{CREffect1,ref-pp-highmult9,ref-pp-highmult10,ref-pp-highmult11}). In order to understand the sensitivity of $\rho(r)$ and $z^{\rm ch}$ on the nature of parton, the gluon-initiated jets and their contributions to the total inclusive jets are estimated and associated effects on these observables are studied. The effect of multiparton interactions are also studied exclusively by using event samples with different number of MPIs. This study will help us to identify and understand the impacts of MPI, CR and gluonic contribution on $\rho(r)$ and $z^{\rm ch}$ distributions.
\par
The paper is organised as follows. The details of the analysis and PYTHIA~8 simulation are discussed in Sec.~\ref{sec:Pythia}. Section~\ref{sec:Obls} describes the observable used in this study and the underlying event estimation technique is detailed in Sec.~\ref{sec:UE}. The obtained results in this work are discussed in Sec.~\ref{sec:Res}. Finally, we conclude with a summary in Sec.~\ref{sec:Summary}. 

\section{\label{sec:Pythia}Pythia Simulation and Analysis Details}
This study is performed on events generated using PYTHIA~8 (version 8.219) Monash~2013~\cite{ref-Pythia8Monash}.
PYTHIA~8 is a multiparton interaction-based pQCD-inspired Monte Carlo event generator widely used for hadronic collisions. It performs transverse-momentum-ordered ($p_{\rm T}$-ordered) parton showering which interleaves the entire perturbative evolution (initial state radiation (ISR), final state radiation (FSR) and MPI) and the angular ordering is imposed by an additional veto~\cite{pythiaShower,InterleavedTuning}. It also includes color reconnection mechanism~\cite{ColorReconnection1,ColorReconnection2}. Hadronization in PYTHIA~8 proceeds via string breaking as described by the Lund string model~\cite{LundStringModel}. The Monash~2013 tune is based on a large set of LHC distributions, starting from a careful comparison and tuning to LEP data. The PDF used is the NNPDF2.3~\cite{NNPDF2pt3} QCD+QED LO with $\alpha_s (M_Z) = 0.130$.
\par
\begin{figure}[!h]
  \centering
  \includegraphics[width=.7\columnwidth]{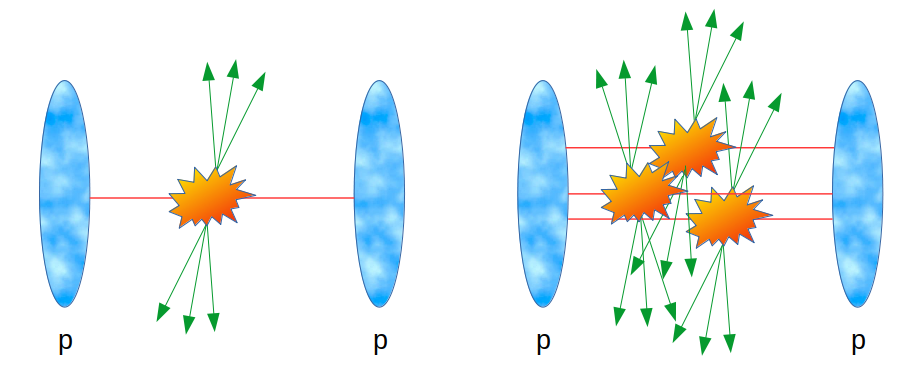}
  \caption{Single hard scattering (left) and multiple hard scatterings, i.e., MPI (right), occurring in a single proton-proton collision.}
  \label{SPS_MPI}
\end{figure}

MPI is a natural consequence of the composite structure of the colliding hadrons, leading to several parton-parton interactions occurring in one hadron-hadron collision (the schematic is shown in Fig.~\ref{SPS_MPI}) and is implemented in PYTHIA~8 through a single unified framework~\cite{MPIinPYTHIA8} that incorporates both soft and hard QCD MPI processes~\cite{PYTHIA8pt2}.

\par
\begin{figure}[!ht]
	\centering
	\begin{subfigure}{.28\textwidth}
		\includegraphics[width=\columnwidth]{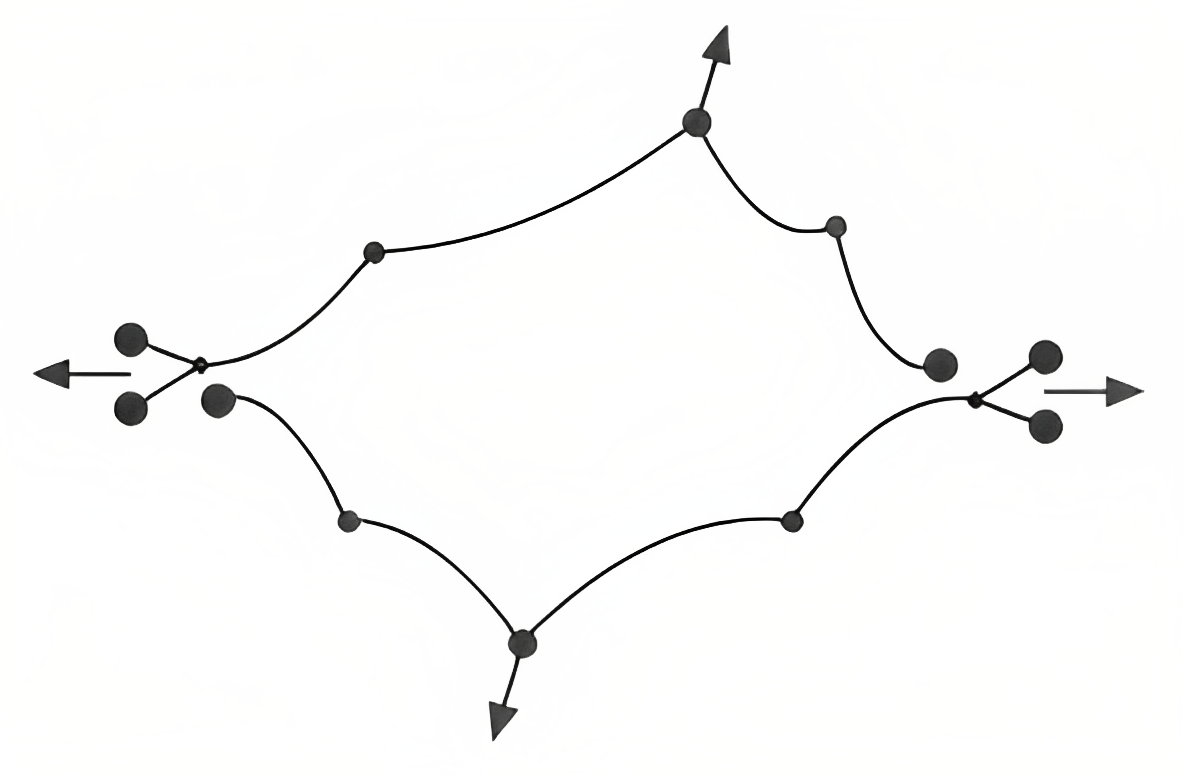}
		\caption{}
		\label{CR-1}
	\end{subfigure}
	\begin{subfigure}{.25\textwidth}
		\includegraphics[width=\columnwidth]{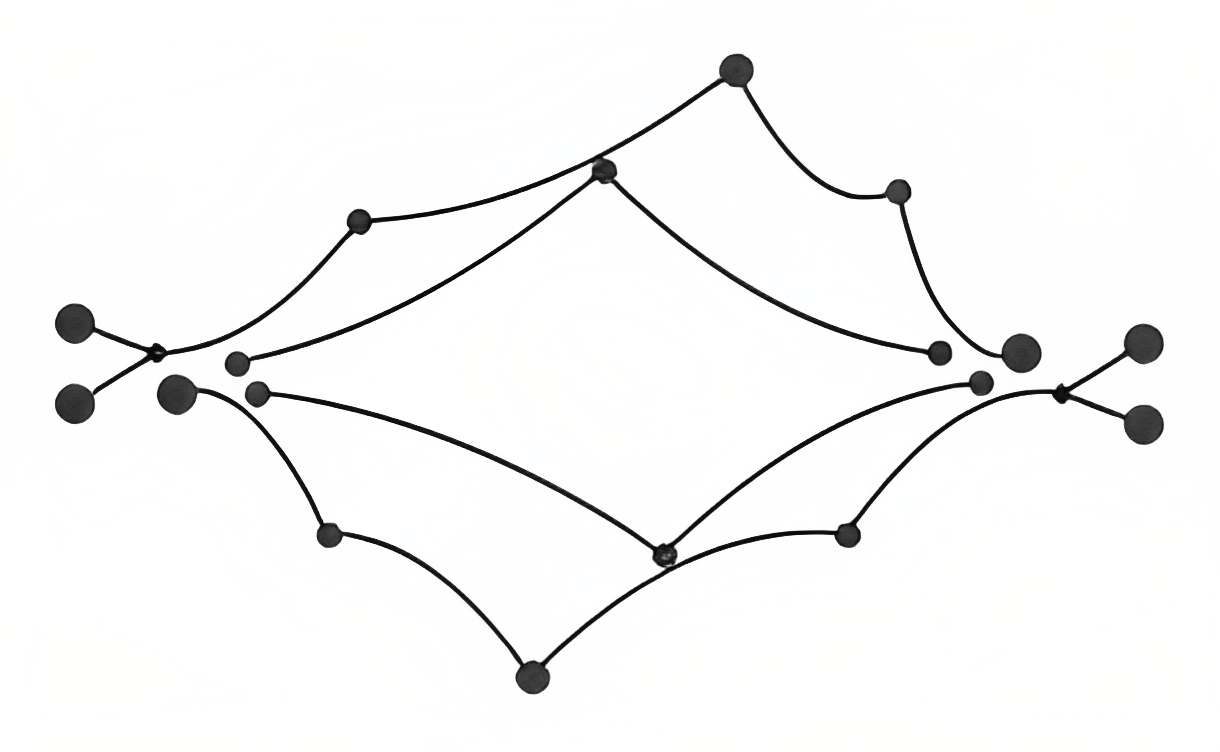}
		\caption{}
		\label{CR-2}
	\end{subfigure}
	\begin{subfigure}{.25\textwidth}
		\includegraphics[width=\columnwidth]{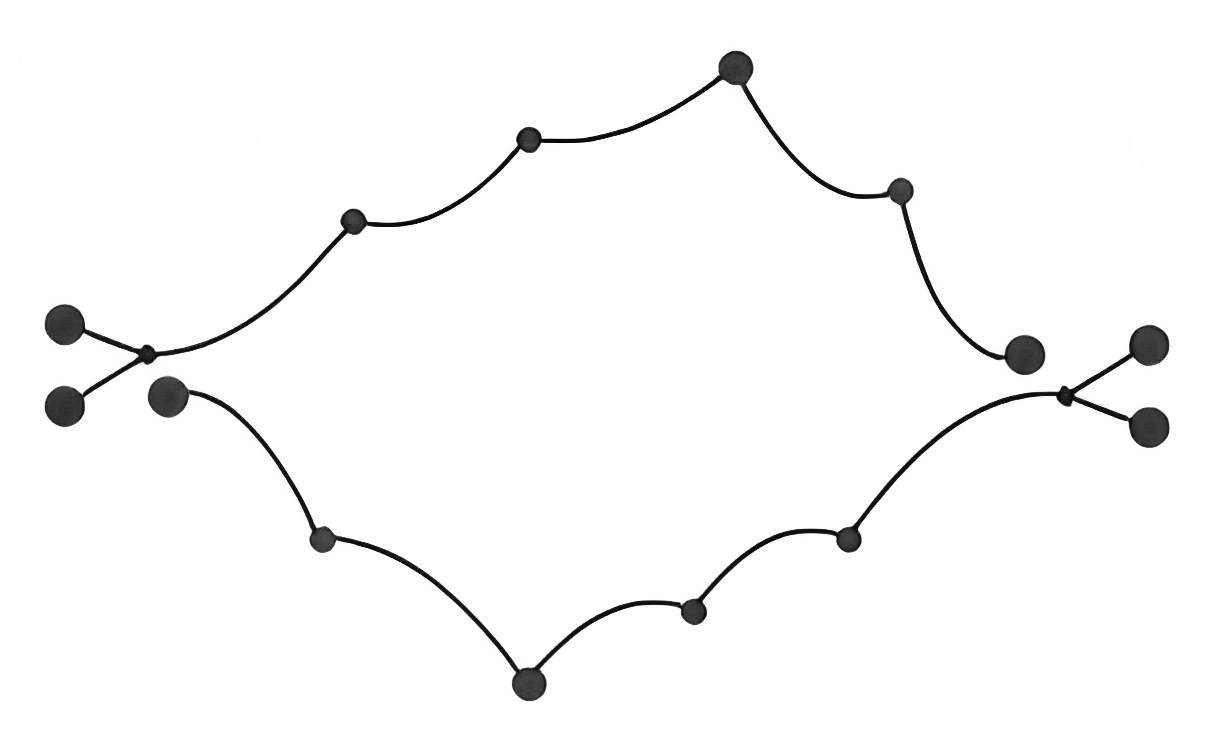}
		\caption{}
		\label{CR-3}
	\end{subfigure}
	\caption{Illustration of the color reconnection mechanism in PYTHIA~8 (image directly extracted from Ref.~\cite{Gustafson:2009qz}). (a) The outgoing gluons are color connected to the projectile and target beam remnants. (b) A second hard scattering with two new strings connected to the beam remnants. (c) Color reconnected partons minimizing the total string length.}
	\label{CRwithMPI}
\end{figure}

The implementation of the color reconnection mechanism in PYTHIA~8 is schematically illustrated in Fig.~\ref{CRwithMPI}. The connection between the outgoing partons and the beam remnants through color strings in case of a single hard scattering is shown in Fig.~\ref{CR-1}. A second hard scattering (Fig.~\ref{CR-2}) can be naively expected to give rise to two new strings connected to the beam remnants. This would result in a proportional increase in multiplicity; however, to successfully fit the data (see Ref.~\cite{Gustafson:2009qz} and references therein) it is instead assumed that the partons are color reconnected so that the total string length gets minimized (Fig.~\ref{CR-3}).

\par
MPIs occurring in a hadronic collision lead to the creation of an environment having several high-momentum partons along with the soft ones in a small region (the overlap area of the colliding hadrons), leading to high-multiplicity events. The evolution of the scattered outgoing partons to final-state collimated hadrons (jets) via fragmentation and hadronization in such an environment is expected to be different compared to the situation with no MPI (only one hard scattering per hadronic collision), which could affect the differential jet shape properties. The fragmentation of independent hard scatterings (MPIs) becomes correlated due to the color reconnection mechanism~\cite{ref-pp-highmult9} described earlier and is, therefore, expected to further modify the differential shape jet properties.
\par
For this study, about 1000 million minimum bias (MB) inelastic events are generated for pp collisions at $\sqrt{s}$~=~13~TeV using PYTHIA~8 Monash~2013 for each of the three following configurations:
\begin{itemize}
\item MPI:~OFF, CR:~OFF -- In this configuration, both multiparton interaction and color reconnection mechanisms of PYTHIA~8 are absent in the process of event generation
\item MPI:~ON, CR:~OFF -- The multiparton interactions are present; however, the color reconnection mechanism is absent in the simulation process in this configuration
\item MPI:~ON, CR:~ON -- Events are generated in presence of both MPI and CR mechanisms of PYTHIA~8
\end{itemize}
The configuration `MPI:~OFF, CR:~ON' is not that important for this study as the effect of color reconnection on the number of produced particles is small in absence of MPI. High-multiplicity (HM) event class is selected as the one that contains 5\% of the total events with the highest multiplicities, based on the number of charged particles produced in the pseudorapidity regions 2.8~$\textless~\eta~\textless$~5.1 and $-\rm{3.7}~\textless~\eta~\textless~-$1.7. The above selection corresponds to the number of charged particles to be greater than 24, 127 and 83 for `MPI:~OFF, CR:~OFF', `MPI:~ON, CR:~OFF' and `MPI:~ON, CR:~ON' configurations respectively. The choice of pseudorapidity range is guided by the experimentally available coverage range of forward detectors~\cite{ref-centrality-v0} used for multiplicity selection. The current study is performed using the particles produced at mid-rapidity to avoid autocorrelation. The generated particles are subjected to the kinematic cuts: $|\eta_{\rm particle}^{\rm ch}|$~\textless~0.9 and $p_{\rm T,\,particle}^{\rm ch}$~\textgreater~0.15~GeV/$c$. The restriction on $\eta$ has given access to the particles only in the mid-rapidity region, whereas the condition for the particles to have $p_{\rm T}$ as low as 0.15~GeV/$c$ has the significance of allowing us to test perturbative and non-perturbative aspects of jet production. The particle selection criteria are chosen to match experimental conditions~\cite{ALICEjetPP7TeV,ALICEjetPP7TeV2}.
\par
Jets are reconstructed with charged particles using an infrared- and collinear-safe sequential recombination anti-$k_{\rm T}$ algorithm~\cite{antikT} from the FastJet package~\cite{FastJet} with jet resolution parameter $R$~(= $\sqrt{(\Delta \eta)^2 + (\Delta \varphi)^2}$)~=~0.4. Jets are accepted for the study if $|\eta_{\rm jet}^{\rm ch}|$~\textless~0.5 (0.9~-~$R$) and jet transverse momentum $p_{\rm T,\,jet}^{\rm ch}$~\textgreater~10~GeV/$c$.
\par
To select gluon-initiated jets, one needs to properly match the hard-scattered partons in an event to the reconstructed charged-particle jets using some effective algorithm. We have taken into consideration an algorithm based on the ``distance of closest approach''. In this algorithm, first we identify the two outgoing hard scattered partons and their flavors (quark or gluon) using the information from PYTHIA event output. In the same event, we then determine unique pairs between these initial hard-scattered partons and reconstructed jets in such a way that the geometrically closest jet is matched to the parent parton. The jets having $p_{\rm T}$ less than 20\% of the matched parton $p_{\rm T}$, are rejected to avoid fake jets. 
This cutoff of 20\% for rejecting fake jets is also varied up to 50\% and no significant change in the final results is observed. The reconstructed jets matched to parent gluons are considered gluon-initiated jets.

\section{\label{sec:Obls}Observables}
In this work, we study the differential transverse momentum
distribution of charged-particle jets ($\frac{1}{N_{\rm events}}\frac{d^2N}{dp_{\rm T}d\eta}$; where $N_{\rm events}$ is the number of events and $N$ is the number of jets), differential jet shape ($\rho(r)$) and jet fragmentation function ($z^{\rm ch}$) for charged-particle jets.
\par
The differential jet shape is related to the radial distribution of jet transverse momentum density inside the jet cone about the jet axis and is defined as:
\begin{equation}
  \rho(r) = \frac{1}{\Delta r} \frac{1}{N_{\rm jets}}\sum_{i=1}^{N_{\rm jets}}p_{\rm T}^i(r - \Delta r/2, r + \Delta r/2)/p_{\rm T,\,jet}^{\rm ch}
\end{equation}
where $r$ is the distance from the jet axis and $p_{\rm T}^i(r - \Delta r/2, r + \Delta r/2)$ denotes summed $p_{\rm T}$ of all particles of $i$-th jet, inside the annular ring between $r - \Delta r/2$ and $r + \Delta r/2$.
\par
The jet fragmentation function represents the fraction of the jet transverse momentum carried by the constituent charged particles and is sensitive to the details of the parton showering process. It is defined as:
\begin{equation}
  z^{\rm ch} = \frac{p_{\rm T,\,particle}^{\rm ch}}{p_{\rm T,\,jet}^{\rm ch}}
\end{equation}
\par
The study of $\rho(r)$ and $z^{\rm ch}$ are very important since these observables are sensitive to both the fragmentation process and the nature of the initial hard-scattered partons (quark or gluon)~\cite{OPALqVSgJet,JetShapesinHadronColliders,JetsinHadronColliders,Vitev:RhoImport,JetShapesImportance}.
In presence of medium, jet constituents lose their energy via inelastic and elastic scatterings due to jet-medium interaction and their mean opening angle also becomes larger. This results in the steepening of $z^{\rm ch}$ and flattening of $\rho(r)$ distributions leading to broadening and softening of jets in medium compared to those in vacuum. The observables $\rho(r)$ and $z^{\rm ch}$ are sensitive to the degree of jet modification in heavy-ion collisions. In case of high-multiplicity pp and p--Pb collisions, these observables might be potentially capable of verifying the conjecture of medium formation in small collision systems~\cite{Vitev:RhoImport,JetShapesImportance}.

\section{\label{sec:UE}Underlying event}
Reconstructed jets are contaminated by the underlying event (UE) which can be defined as the charged particles except those coming from the fragmentation of hard-scattered partons. The UE mostly consists of particles from the beam-beam remnants, initial and final state radiations and contributions from MPIs~\cite{CDF}. The empirical models used for the description of the non-perturbative aspects in the evolution of a high-energy scattering event do not allow to clearly distinguish particles originating from hard processes and the underlying event~\cite{UEpp13TeVALICE}. The UE is estimated from circular regions transverse to the measured jet cones, known as perpendicular cone (PC) method, in each event. The size of these circular regions is defined by the radius of the jet $R$~=~0.4 at the same pseudorapidity as the considered jet but offset at an azimuthal angle $\Delta \varphi = \pm \pi/2$ relative to the jet axis. For the estimation of UE contribution to the $\rho(r)$ distribution, annular rings of the same size as those inside the jet cones are considered inside each of the perpendicular circular regions. The UE contributions to the $\rho(r)$ and $z^{\rm ch}$ distributions are respectively calculated using the following expressions:
\begin{equation}
  \rho^{\rm UE}(r) = \frac{1}{\Delta r} \frac{1}{N_{\rm PC}}\sum_{i=1}^{N_{\rm PC}}p_{\rm T}^i(r - \Delta r/2, r + \Delta r/2)/p_{\rm T,\,jet}^{\rm ch}
\end{equation}
where $N_{\rm PC}$ is the number of perpendicular cones, $r$ is the distance from the axis of the perpendicular cone and $p_{\rm T}^i(r - \Delta r/2, r + \Delta r/2)$ denotes summed $p_{\rm T}$ of all particles inside the annular ring between $r - \Delta r/2$ and $r + \Delta r/2$ and
\begin{equation}
	z^{\rm ch, UE} = \frac{p_{\rm T,\,particle}^{\rm ch,\,PC}}{p_{\rm T,\,jet}}
\end{equation}
where $p_{\rm T,\,particle}^{\rm ch,\,PC}$ is the $p_{\rm T}$ of particle in the perpendicular cone. The subtraction of UE is performed on a statistical basis to obtain the corrected distributions.
\par
In order to perform a systematic check, a random cone (RC) method is also applied to estimate the underlying event contribution in the studied jet observables. In this method, two cones are randomly generated with the same $\eta$ as the considered jet, but in the expanded transverse region (one cone within $\pi/3 < \Delta \varphi < 2\pi/3$ and another within $-\pi/3 < \Delta \varphi < -2\pi/3$, $\Delta \phi$ being the difference between the considered jet and the random cone in azimuthal angle $\varphi$) with respect to the jet axis, unlike at a fixed value of azimuthal angle ($\Delta \varphi = \pm \pi/2$) as in PC method. No significant difference is observed in the UE contribution as compared to PC method.

\section{\label{sec:Res}Results and Discussion}
\subsection{\label{subsec:DataComp}Comparison with experimental data}
\begin{figure}[!ht]
	\centering
	\includegraphics[scale=.45]{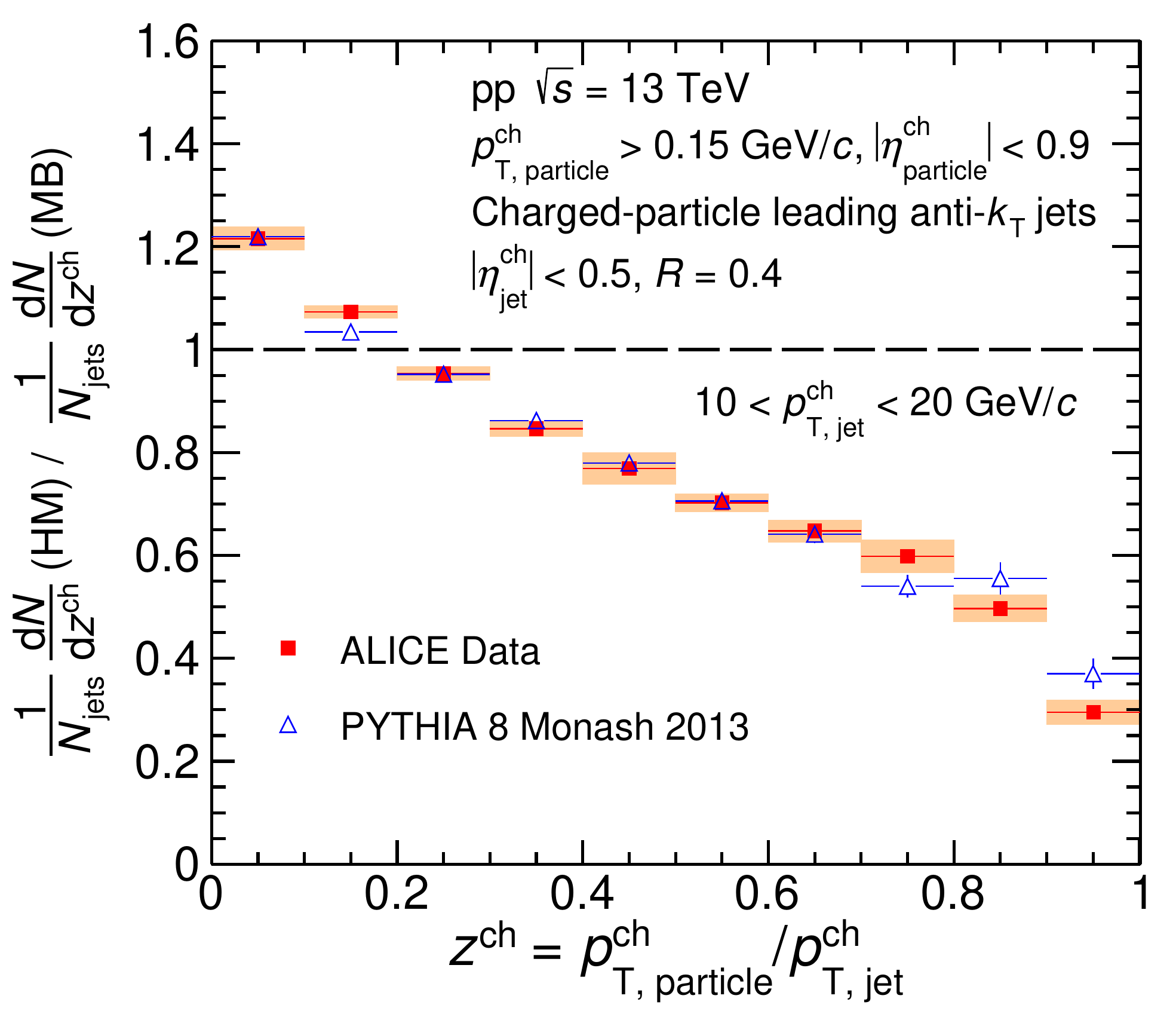}
	\caption{Ratio of charged-particle jet fragmentation distributions for leading jets in the interval 10~$<p_{\rm T,\,jet}^{\rm ch}<$~20~GeV/$c$ between high-multiplicity and minimum bias event classes in pp collisions at $\sqrt{s}$ = 13 TeV. The red solid boxes and blue open triangles represent ALICE data and PYTHIA 8 Monash 2013 predictions respectively. The systematic uncertainty in data is represented by orange bands.}
	\label{FFLeadingJets10to20}
\end{figure}

First we compare the ALICE data~\cite{JetFragPreliDB} to PYTHIA~8 predictions in Fig.~\ref{FFLeadingJets10to20} which shows the ratio of charged-particle jet fragmentation distributions for leading jets (jet with the highest $p_{\rm T}$ in an event) in the interval 10~$<p_{\rm T,\,jet}^{\rm ch}<$~20~GeV/$c$ between high-multiplicity and minimum bias event classes in pp collisions at $\sqrt{s}$ = 13 TeV. 
Similar track and jet selection criteria as in data are applied in PYTHIA~8 for this comparison. However the event selection is slightly different. In Ref.~\cite{JetFragPreliDB}, HM events are selected using the detector level information (based on the two scintillator detectors V0A and V0C in ALICE) both for data and PYTHIA 8 simulation whereas in this study we use the information of final-state particles in the same pseudorapidity coverage as described in Sec.~\ref{sec:Pythia}.
In this case, the selected HM event class comprises of 0.1\% of the total events with the highest multiplicities to match the selection criteria of Ref.~\cite{JetFragPreliDB}.
Nonetheless, it is found that, even without implementation of any medium effects, PYTHIA~8 reproduces the jet modification observed in data fairly well. This observation is very interesting and indicates that PYTHIA~8 indeed incorporates underlying physics mechanism(s) which can capture the features of jet modification observed in experimental data. It is therefore very important to understand these underlying physics mechanism(s) in PYTHIA~8. We investigate the effect of MPI and CR on the transverse momentum ($p_{\rm T}$) spectra of the charged-particle jets, jet shape ($\rho(r)$) and jet fragmentation function ($z^{\rm ch}$) in MB and HM event classes.
\subsection{\label{subsec:ResJetPt}Jet $p_{\rm T}$ spectra}
\begin{figure}[!h]
  \centering
  \includegraphics[scale=.45]{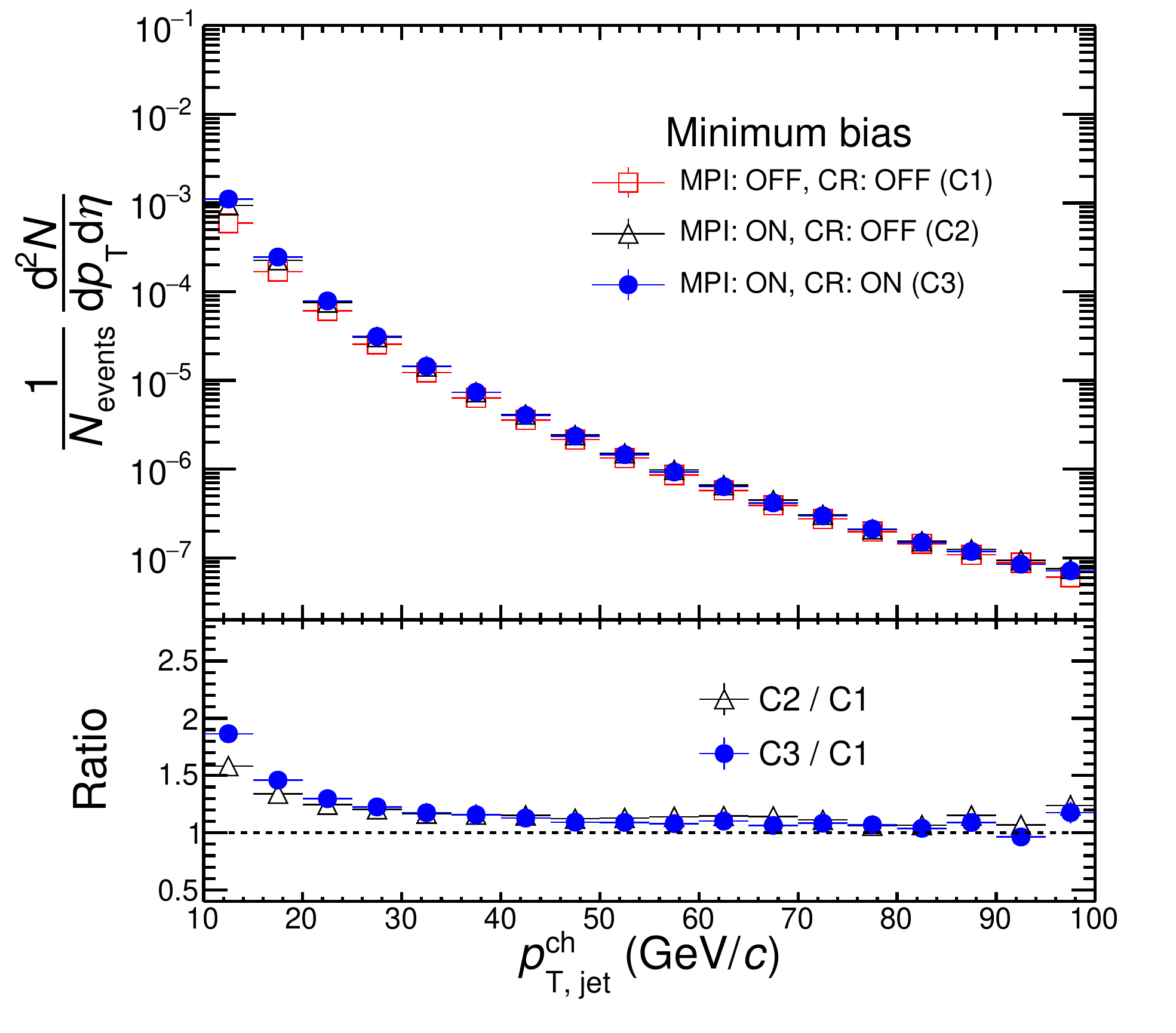}
  \caption{Top panel: Inclusive charged-particle jet $p_{\rm T}$ spectra in pp collisions at $\sqrt{s}$~=~13~TeV using PYTHIA~8 Monash~2013 for MB event class. Open red boxes, open black triangles and solid blue circles correspond to `MPI:~OFF, CR:~OFF' (C1), `MPI:~ON, CR:~OFF' (C2) and `MPI:~ON, CR:~ON' (C3) configurations respectively. Bottom panel: Ratios of $p_{\rm T}$ spectra for the last two configurations (C2 and C3) with respect to the first configuration (C1).}
  \label{JetPtSpectra}
\end{figure}

We compare the $p_{\rm T}$ spectra of charged-particle jets between `MPI:~ON, CR:~ON', `MPI:~ON, CR:~OFF' and `MPI:~OFF, CR:~OFF' configurations in Fig.~\ref{JetPtSpectra} for MB event class.
The top panel shows the $p_{\rm T}$ spectra of charged-particle jets and the bottom panel shows the ratios of spectra in `MPI:~ON, CR:~ON' and `MPI:~ON, CR:~OFF' configurations with that in `MPI:~OFF, CR:~OFF' configuration. It is interesting to observe that the rate of jet production increases when MPI effects are switched ON. Compared to `MPI:~OFF, CR:~OFF' configuration, the rate of jet production increases by about 60\% when only MPI is switched ON while it increases to about 86\% when CR is switched ON in addition to MPI at low jet $p_{\rm T}$ (10 - 15 GeV/$c$). The observed increase in the rate of jet production decreases with increasing jet $p_{\rm T}$ for both the configurations. This is an indication of multiple jet production due to multiparton interactions which is expected to be more prominent at low jet $p_{\rm T}$~\cite{PYTHIA8pt2}.
\subsection{\label{sec:ResJetShapeAndFragmentation}Jet shape and jet fragmentation}
\begin{figure}[!h]
  \centering
  \begin{subfigure}[b]{.49\textwidth}    
    \includegraphics[width=\columnwidth]{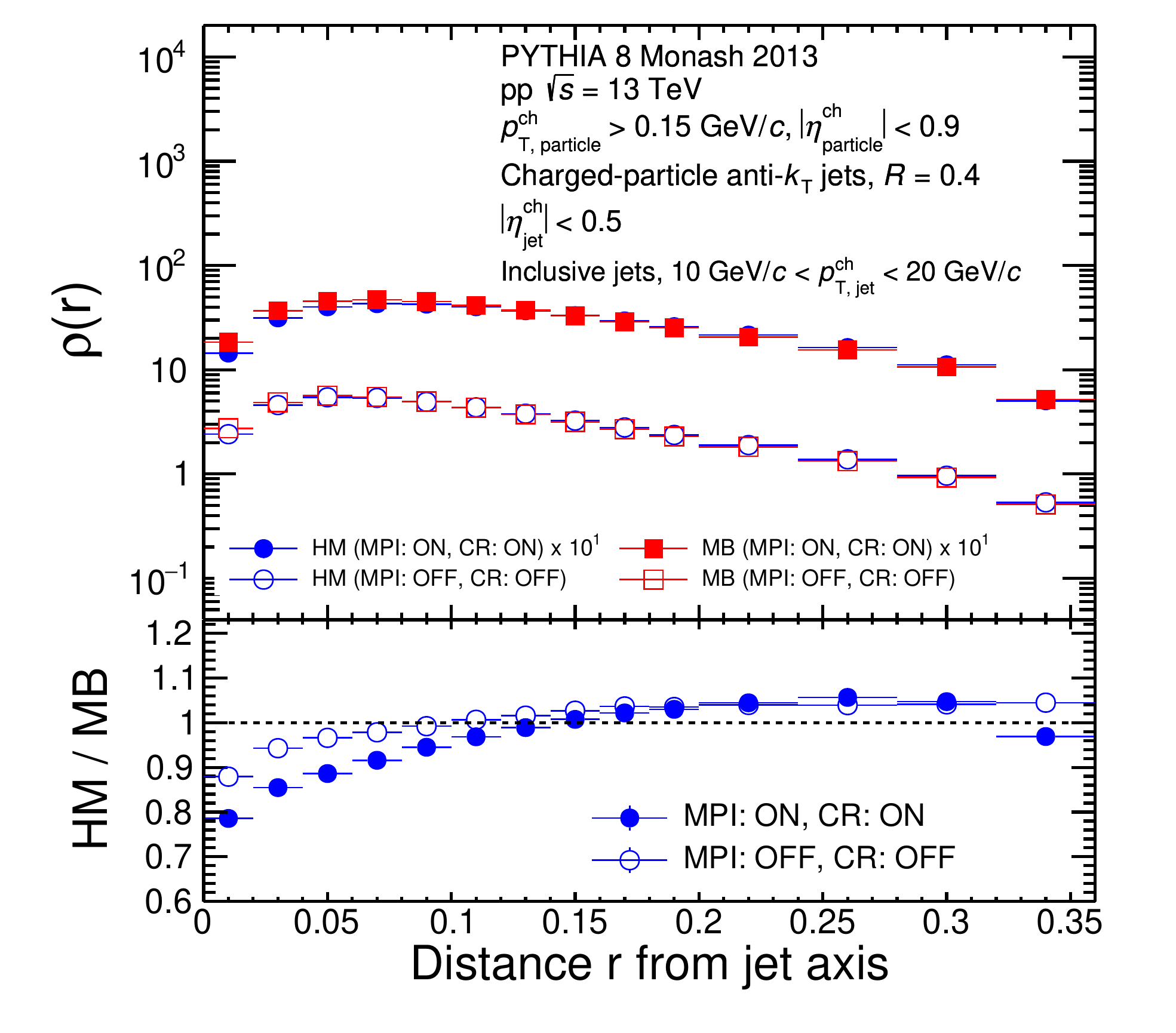}
    \caption{}
    \label{rhoInclJets10to20}
  \end{subfigure}
  \begin{subfigure}[b]{.49\textwidth}
    \includegraphics[width=\columnwidth]{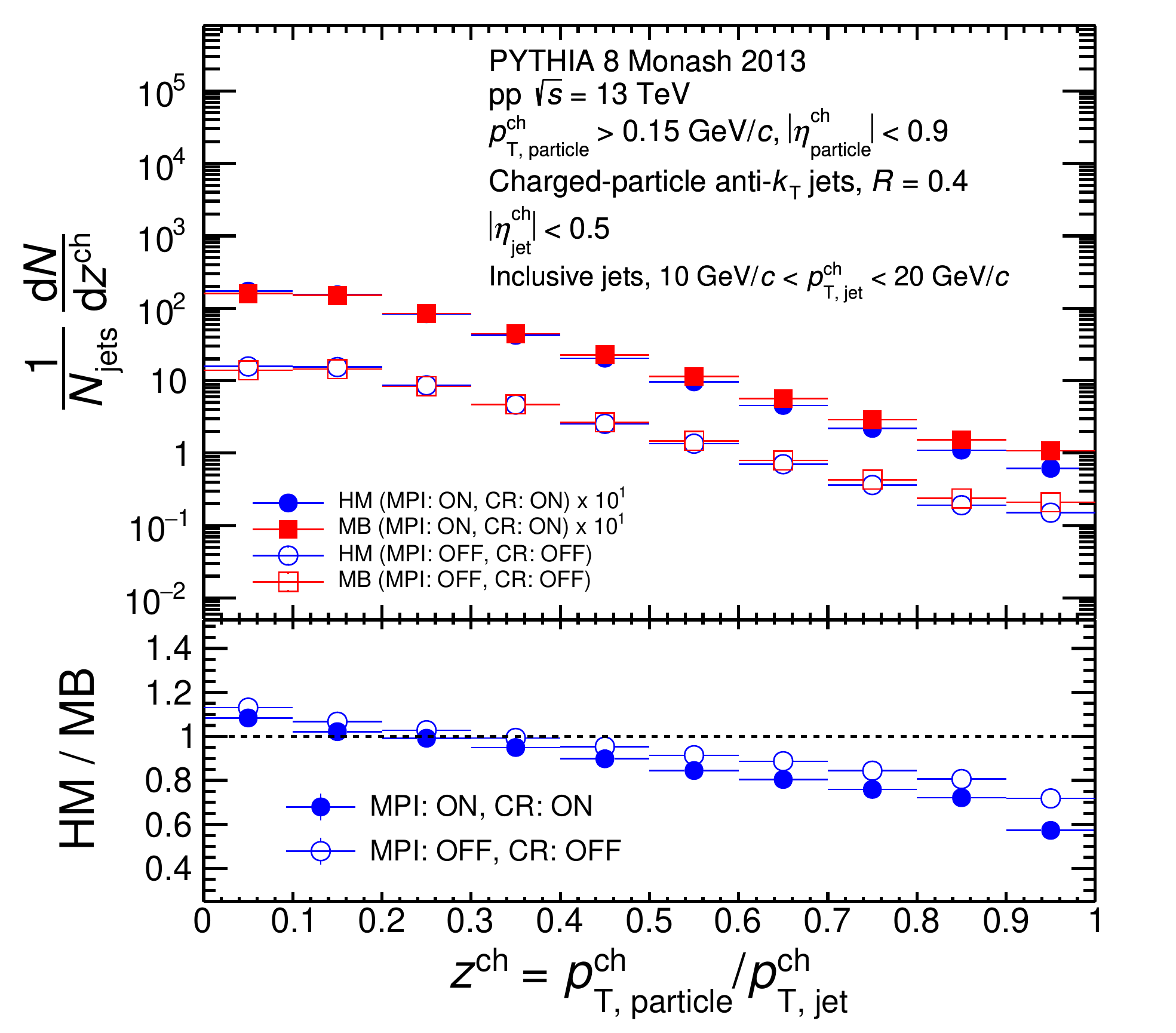}
    \caption{}
    \label{FFInclJets10to20}
  \end{subfigure}
  \caption{Top panels: Distributions of (a) $\rho(r)$ and (b) $z^{\rm ch}$ for inclusive charged-particle jets in pp collisions at $\sqrt{s}$~=~13~TeV using PYTHIA~8 Monash~2013 in the interval 10~$<p_{\rm T,\,jet}^{\rm ch}<$~20~GeV/$c$ for `MPI:~ON, CR:~ON' and `MPI:~OFF, CR:~OFF' configurations. Blue circles and red boxes correspond to HM and MB event classes respectively. Bottom panels: Ratios of (a) $\rho(r)$ and (b) $z^{\rm ch}$ distributions between HM and MB event classes.}
  \label{InclJets10to20}
\end{figure}
Figures~\ref{rhoInclJets10to20} and \ref{FFInclJets10to20} show respectively the jet shape ($\rho(r)$) plotted as a function of distance $r$ from the jet axis and jet fragmentation ($z^{\rm ch}$) for inclusive charged-particle jets in the interval 10~$<p_{\rm T,\,jet}^{\rm ch}<$~20~GeV/$c$ for HM and MB event classes in the top panels and their corresponding ratios in the bottom panels. The solid and open markers represent `MPI:~ON, CR:~ON' and `MPI:~OFF, CR:~OFF' configurations respectively. Solid markers are scaled by a factor of 10 for better readability. Significant amount of modifications in jet shape and jet fragmentation function are observed in HM event class compared to that in MB event class.

The modification in jet shape is more prominent when CR and MPI effects are switched ON. As evident from the ratio plots of $\rho(r)$ as shown in Fig.~\ref{rhoInclJets10to20} (bottom panel), the core of the jet in HM event class is depleted by about 22\% and the energy is redistributed away from the jet axis resulting in enhancement in $\rho(r)$ at larger $r$~($>0.15$). Results are shown only up to $r$~=~0.36 as the last bin ($r$~=~0.36~-~0.4) is significantly affected by underlying event contribution and also by statistical fluctuations. The ratio plots of $z^{\rm ch}$ distributions as shown in Fig.~\ref{FFInclJets10to20} (bottom panel) illustrate similar observation where the production of high $z^{\rm ch}$ particles are substantially suppressed (by about 40\% at $z^{\rm ch}\rightarrow$ 1) in HM event class compared to that in MB. We look at the distributions of the number of MPIs ($N_{\rm MPI}$) in both MB and HM event classes with `MPI:~ON, CR:~ON' configuration and evaluated their mean values using the expression:
\begin{equation}  
  \langle N_{\rm MPI} \rangle = \frac{1}{N_{\rm events}}\sum_{i=1}^{N_{\rm events}} N_{\rm MPI}^{i}
  \label{AvgnMPI}
\end{equation}
where $N_{\rm events}$ is the total number of events in the selected event class and $N_{\rm MPI}^{i}$ is the number of MPIs in the $i$-th event, obtained directly from PYTHIA~8.
\begin{figure}[!h]
  \centering
  \includegraphics[scale=.48]{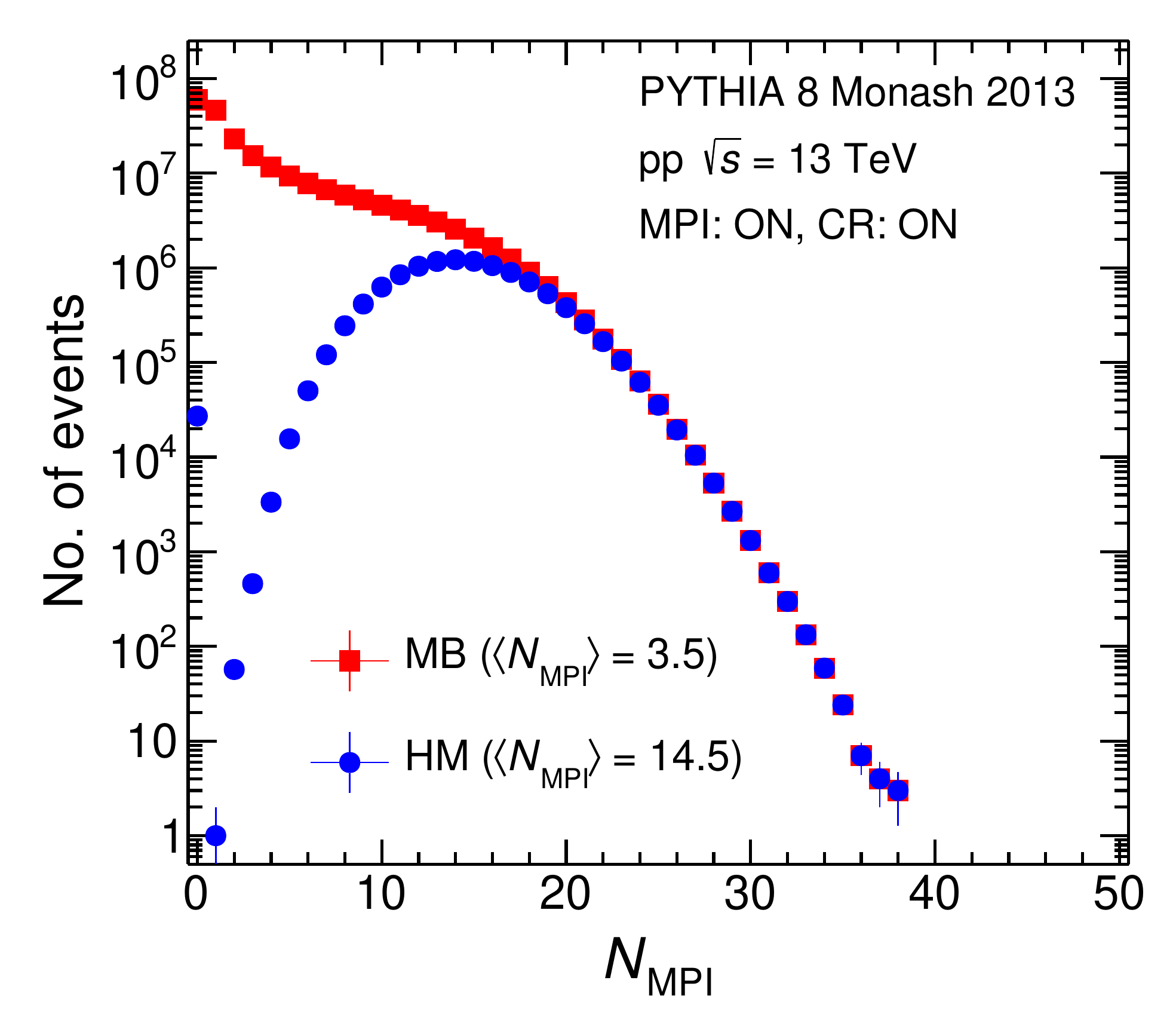}
  \caption{Distributions of no.~of MPIs for HM and MB event classes in `MPI:~ON, CR:~ON' configuration.}
  \label{refMPI}
\end{figure}
Figure~\ref{refMPI} shows the $N_{\rm MPI}$ distributions for the two event classes. It is important to note that the average number of multiparton interactions in HM event class is found to be much larger (14.5) compared to that (3.5) in MB event class.
\par
It is very interesting to notice that the effect of redistribution of energy from the jet core to outer radii and the suppression of high $z^{\rm ch}$ particles are significantly reduced when the CR and MPI effects in PYTHIA~8 are switched OFF. A small depletion of the core and enhancement at larger $r$ for $\rho(r)$ and suppression at high $z^{\rm ch}$ for jet fragmentation distribution are, however, still present. The origin of this residual effect of jet modification in `MPI:~OFF, CR:~OFF' configuration can be understood in terms of difference in the contribution of gluon-initiated jets in HM event class compared to that in MB as discussed below. 
\par
Gluon-initiated jets are expected to fragment into more constituents and become softer and broader compared to quark-initiated jets due to their large color factor~\cite{QvsGjet1,QvsGjet2}. By applying the matching procedure described in Sec.~\ref{sec:Pythia}, we identify the charged-particle jets initiated from the initial hard-scattered partons (quark or gluon) and estimate the gluonic contribution as the fraction of gluon-initiated jets out of the inclusive matched jets. We find that the inclusive matched jets in HM event class contain higher contribution from gluon-initiated jets compared to that in MB event class. For the jets in the interval 10~$<p_{\rm T,\,jet}^{\rm ch}<$~20~GeV/$c$ (40~$<p_{\rm T,\,jet}^{\rm ch}<$~60~GeV/$c$), the contribution from gluon-initiated jets being 86\% (81\%) in HM while 75\% (73\%) in MB event class. 
\begin{figure}[!h]
  \centering
  \begin{subfigure}[b]{.49\textwidth}    
    \includegraphics[width=\columnwidth]{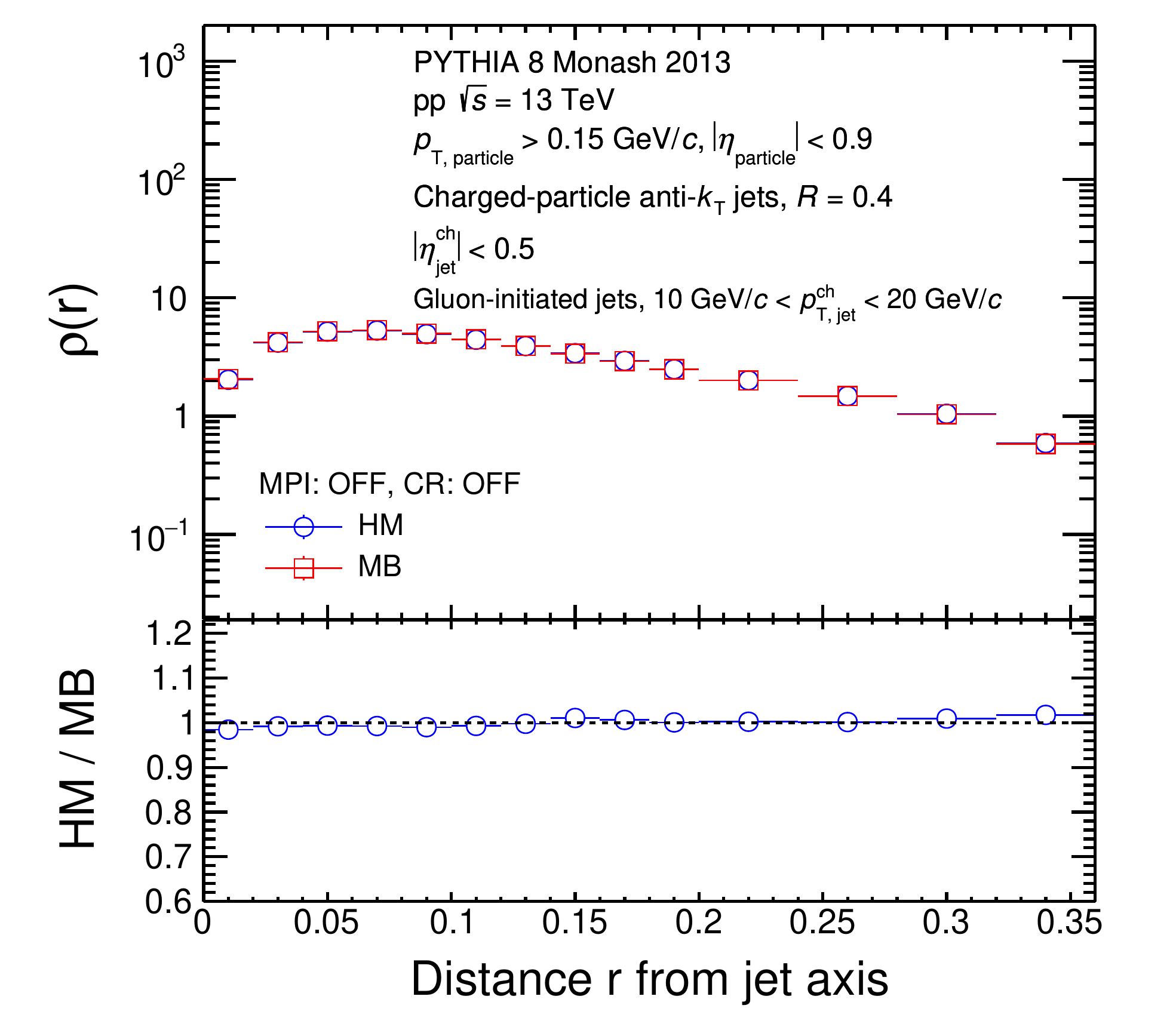}
    \caption{}
    \label{rhoChGJets10to20}
  \end{subfigure}
  \begin{subfigure}[b]{.49\textwidth}
    \includegraphics[width=\columnwidth]{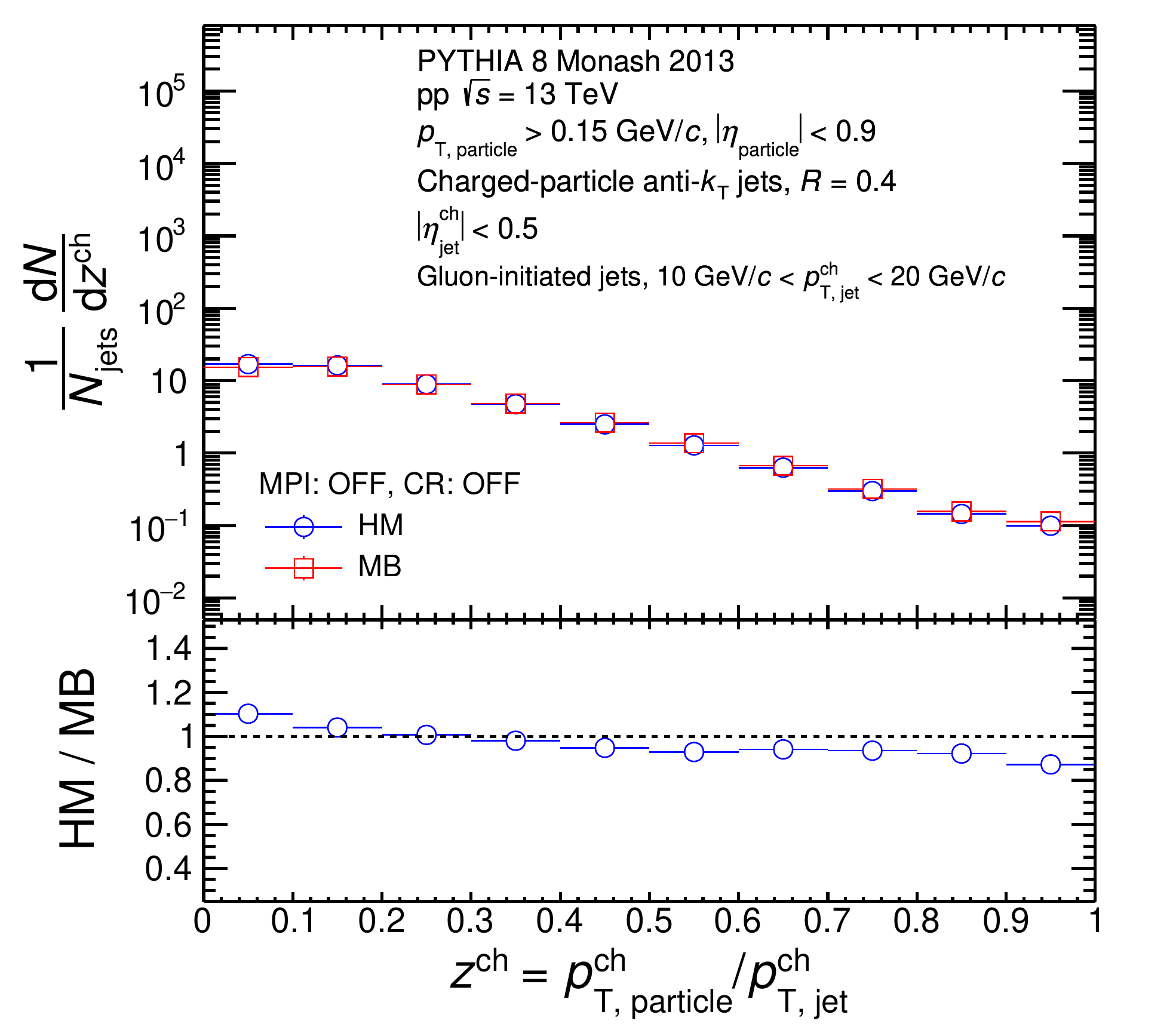}
    \caption{}
    \label{FFChGJets10to20}
  \end{subfigure}  
  \caption{Top panels: Distributions of (a) $\rho(r)$ and (b) $z^{\rm ch}$ for gluon-initiated jets in pp collisions at $\sqrt{s}$~=~13~TeV using PYTHIA~8 Monash~2013 in the interval 10~$<p_{\rm T,\,jet}^{\rm ch}<$~20~GeV/$c$ for `MPI:~OFF, CR:~OFF' configuration. Blue circles and red boxes correspond to HM and MB event classes respectively. Bottom panels: Ratios of (a) $\rho(r)$ and (b) $z^{\rm ch}$ distributions between HM and MB event classes.}
  \label{ChGJets10to20}
\end{figure}
This difference in the gluonic contribution in the HM event class compared to MB is therefore expected to modify the distributions of $\rho(r)$ and $z^{\rm ch}$. To remove this dependence on the difference in gluonic contribution, we look at the ratios of $\rho(r)$ and $z^{\rm ch}$ distributions between HM and MB event classes for gluon-initiated jets only with `MPI:~OFF, CR:~OFF' configuration as shown in Fig.~\ref{ChGJets10to20}. We find that the residual effects observed in case of inclusive jets further get reduced and the ratio tends towards unity.
\par
\begin{figure}[!h]
  \centering
  \begin{subfigure}[b]{.49\textwidth}    
    \includegraphics[width=\columnwidth]{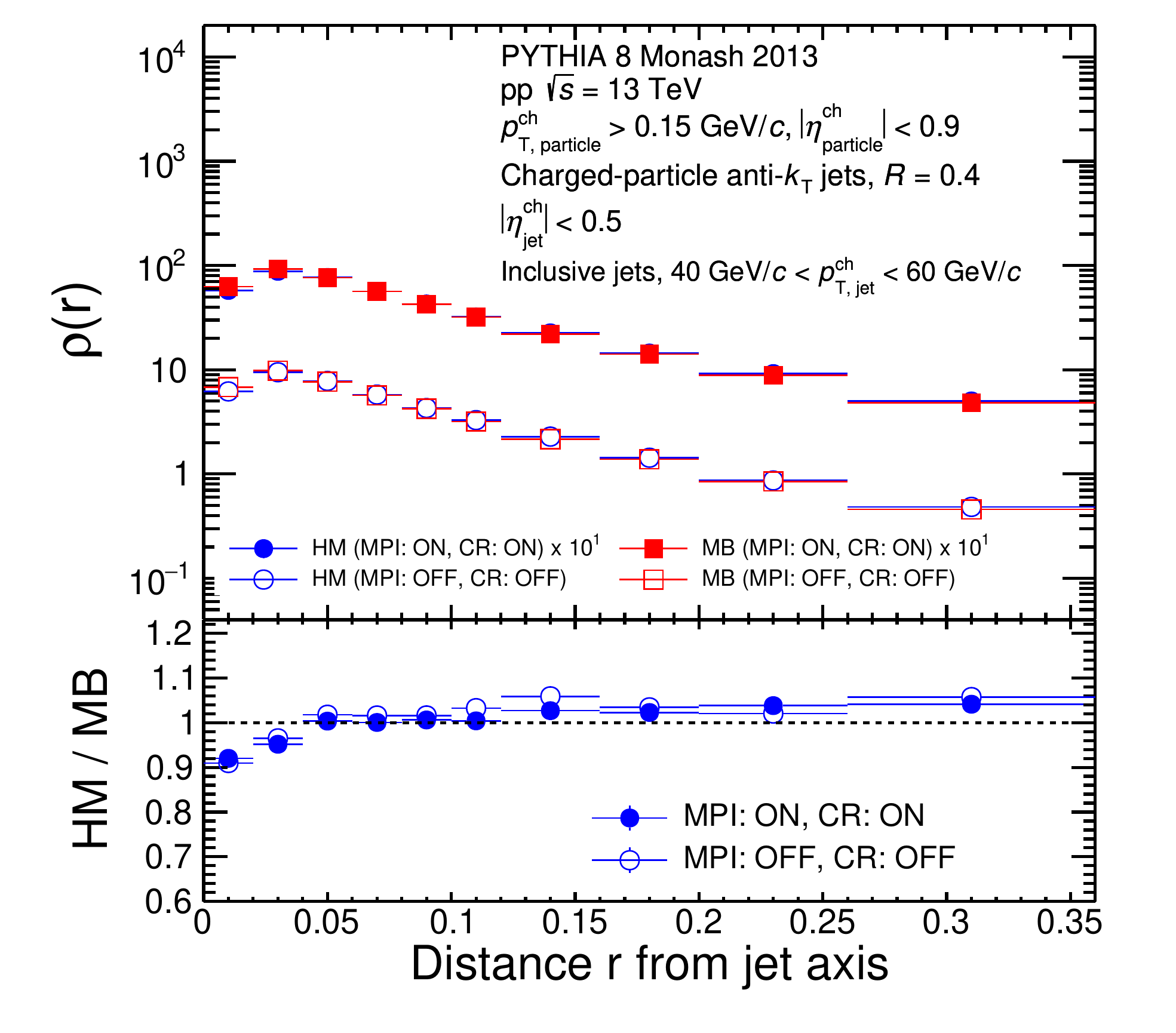}
    \caption{}
    \label{rhoInclJets40to60}
  \end{subfigure}
  \begin{subfigure}[b]{.49\textwidth}
    \includegraphics[width=\columnwidth]{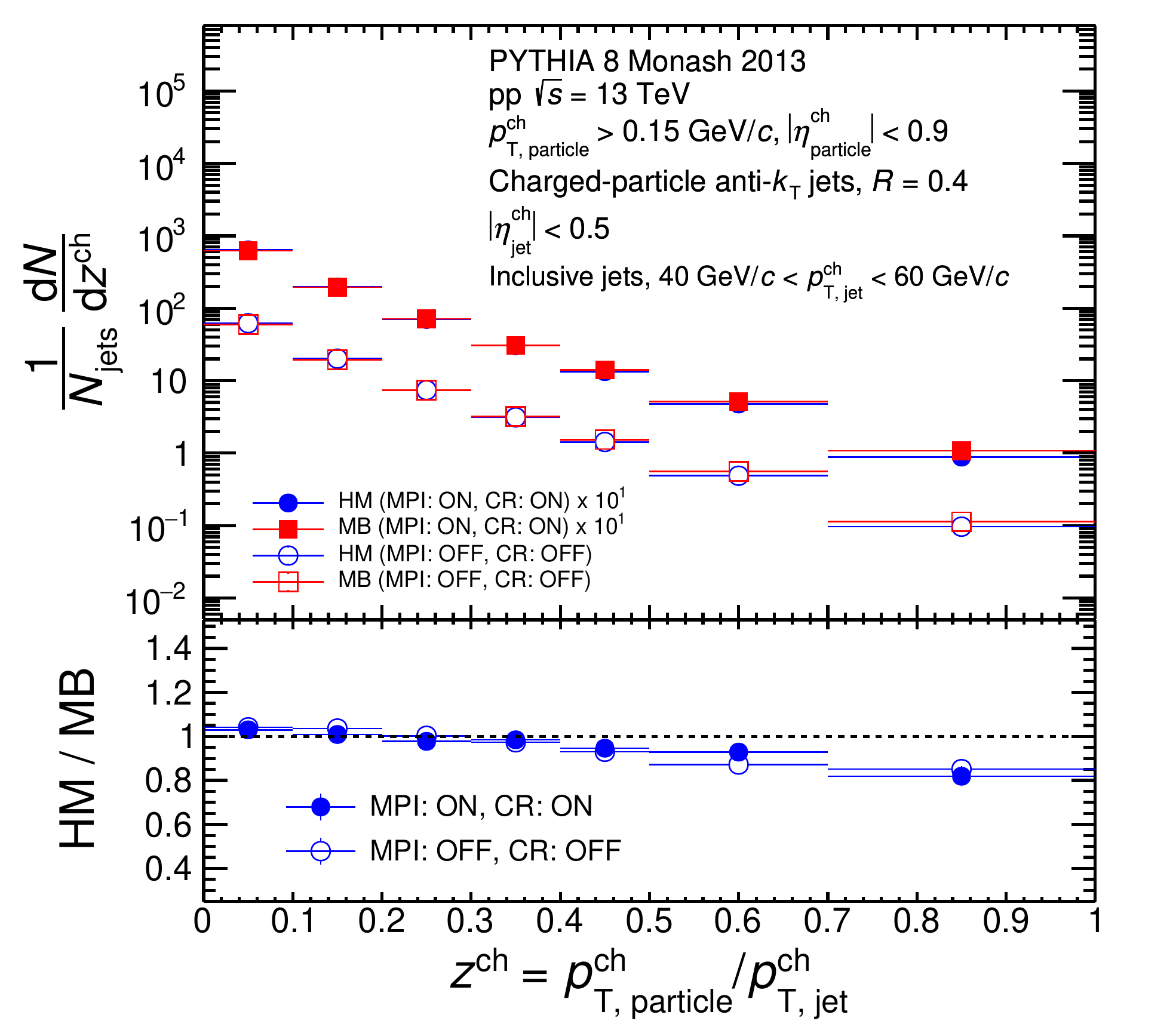}
    \caption{}
    \label{FFInclJets40to60}
  \end{subfigure}
  \caption{Top panels: (a) Inclusive charged-particle jet shape ($\rho(r)$) distributions and (b) jet fragmentation ($z^{\rm ch}$) distributions in pp collisions at $\sqrt{s}$~=~13~TeV using PYTHIA~8 Monash~2013 in the interval 40~$<p_{\rm T,\,jet}^{\rm ch}<$~60~GeV/$c$ for `MPI:~ON, CR:~ON' and `MPI:~OFF, CR:~OFF' configurations. Blue circles and red boxes correspond to HM and MB event classes respectively. Bottom panels: Ratios of (a) $\rho(r)$ and (b) $z^{\rm ch}$ distributions between HM and MB event classes.}  \label{InclJets40to60}
\end{figure}

The effect of MPI and CR is dominant at low $p_{\rm T}$~\cite{PYTHIA8pt2}, therefore the observed jet modification is expected to get reduced at higher jet transverse momentum region. Figures~\ref{rhoInclJets40to60} and~\ref{FFInclJets40to60} show respectively the distributions of $\rho(r)$ and  $z^{\rm ch}$ in the interval 40~$<p_{\rm T,\,jet}^{\rm ch}<$~60~GeV/$c$ for HM and MB event classes in the top panels and their ratios in the bottom panels. It is found that the amount of jet modification is significantly reduced at higher jet $p_{\rm T}$.
\par
We conclude from the above observations (Figs.~\ref{InclJets10to20},~\ref{ChGJets10to20} and~\ref{InclJets40to60}) that the main sources responsible for modifications of $\rho(r)$ and $z^{\rm ch}$ in PYTHIA~8 in HM event class compared to that in MB event class are the mechanisms of MPI and CR and change in the contribution of gluon-initiated jets in HM event class.

In order to further understand the effect of MPI on jet modification more accurately in a quantitative manner, we perform a systematic study using only the jet shape observable $\rho(r)$ with event samples having different number of MPIs as discussed in next subsection. The effect of CR on jet modification is also quantified.

\subsection{\label{subsec:DepMPICR}Effect of MPI and CR on observed jet modification}
To investigate the role of multiparton interactions in modification of $\rho(r)$, we divide the event samples based on the number of MPIs into four different classes I, II, III and IV containing events with the number of MPIs $>$ 4, 8, 12 and 20 respectively. This study is performed using PYTHIA 8 with default configuration (`MPI:~ON, CR:~ON') and for inclusive jets in the interval 10~$<p_{\rm T,\,jet}^{\rm ch}<$~20~GeV/$c$. We estimate the average number of multiparton interactions ($\langle N_{\rm MPI} \rangle$) for different event classes using Eq.~\ref{AvgnMPI} and correlate it to the amount of modification observed in $\rho(r)$. Table~\ref{table:1} shows the values of $\langle N_{\rm MPI} \rangle$ for different event classes. The value of $\langle N_{\rm MPI} \rangle$ increases from 9.1 to 22 while going from event class I to IV.
\begin{table}[h!]
  \caption{Values of $\langle N_{\rm MPI} \rangle$}
  \label{table:1}
  \begin{center}
    \setlength{\tabcolsep}{20pt}
    \renewcommand{\arraystretch}{1.2}
    \begin{tabular}{|c|c|}      
      \hline
      Event class & $\langle N_{\rm MPI} \rangle$ \\
      \hline
      MB & 3.5\\
      \hline
      I & 9.1\\
      \hline
      II & 12.4\\
      \hline
      III & 15.3\\
      \hline
      IV & 22\\
      \hline
    \end{tabular}
  \end{center}	
\end{table}
\begin{figure}[!h]
  \centering
  \includegraphics[scale=.45]{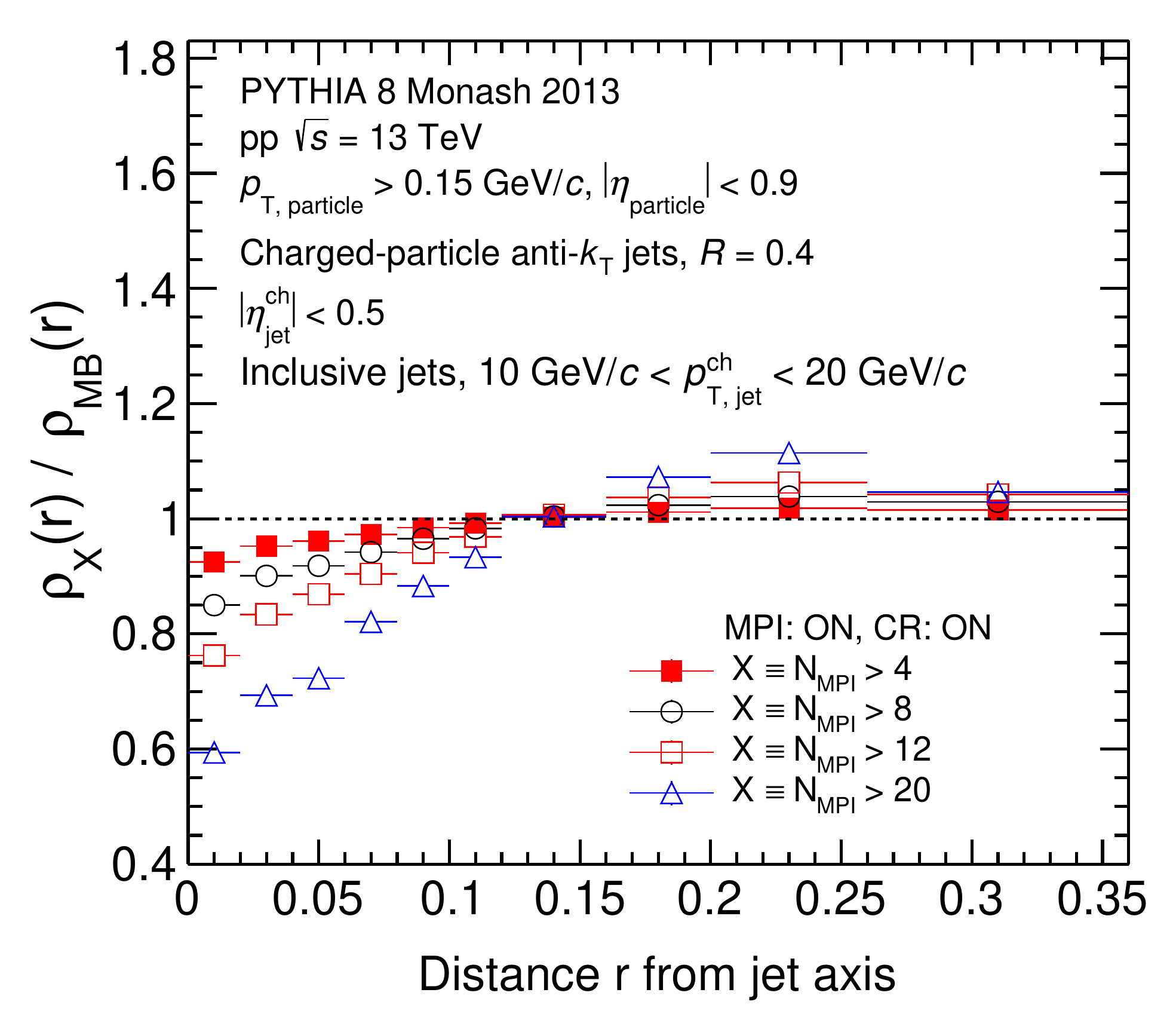}
  \caption{Ratios of $\rho(r)$ distributions of different event classes (I: red solid box, II: black open circle, III: red open box, IV: blue open triangle) with respect to that in MB event class.}
  \label{rhoEffectOfnMPI}
\end{figure}
Figure~\ref{rhoEffectOfnMPI} shows the ratios of $\rho(r)$ distributions of different event classes with respect to that in MB event class. The amount of modification observed in $\rho(r)$ at jet core increases for event classes with larger $\langle N_{\rm MPI} \rangle$. For event class IV where $\langle N_{\rm MPI} \rangle$ = 22, the modification at the jet core is the highest reaching up to 40\%. This observation exclusively supports the existence of a direct connection between the amount of modification of $\rho(r)$ and the number of MPIs in PYTHIA~8.
\par
\begin{figure}[!ht]
  \centering
  \includegraphics[scale=0.45]{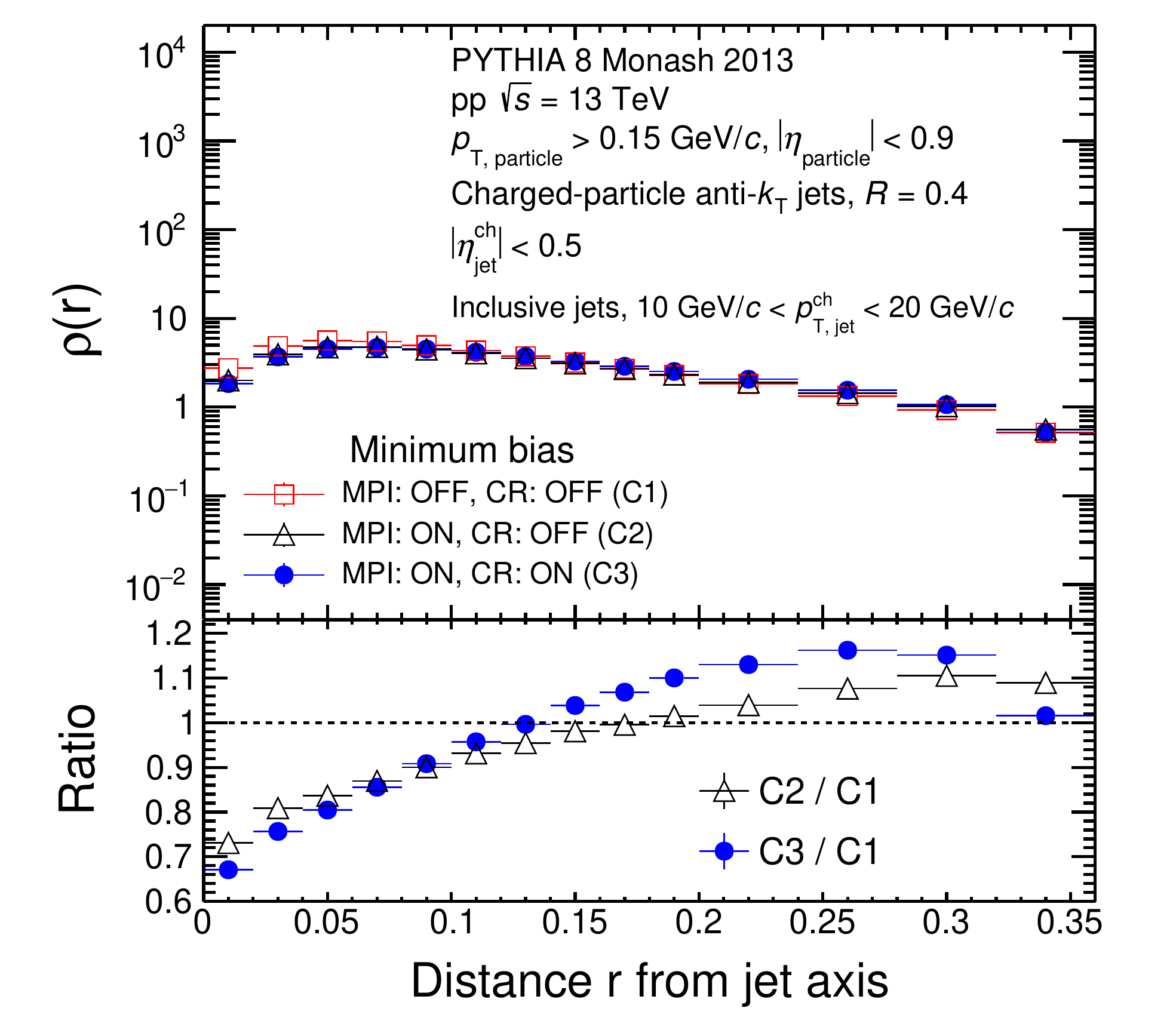}
  \caption{Top panel: Inclusive charged-particle jet shape ($\rho(r)$) distributions in pp collisions at $\sqrt{s}$~=~13~TeV using PYTHIA~8 Monash~2013 in the interval 10~$<p_{\rm T,\,jet}^{\rm ch}<$~20~GeV/$c$ in MB event class. Open red boxes, open black triangles and solid blue circles correspond to `MPI:~OFF, CR:~OFF' (C1), `MPI:~ON, CR:~OFF' (C2) and `MPI:~ON, CR:~ON' (C3) configurations respectively. Bottom panel: Ratios of $\rho(r)$ distributions for the last two configurations (C2 and C3) with respect to the first configuration (C1).}
  \label{refFigRho3}
\end{figure}

To quantify the effect of CR, we compare the $\rho(r)$ distributions for `MPI:~ON, CR:~ON' and `MPI:~ON, CR:~OFF' configurations with `MPI:~OFF, CR:~OFF' configuration in one event class (MB) for inclusive jets in the interval 10~$<p_{\rm T,\,jet}^{\rm ch}<$~20~GeV/$c$ as shown in Fig.~\ref{refFigRho3}.
Compared to `MPI:~OFF, CR:~OFF' configuration, about 26\% (7\%) of modification of jet core, $r$~=~0\,-\,0.02 (outer region, $r$~=~0.24\,-\,0.28) is observed when only MPI is switched ON while the modification increases to 33\% (16\%) when CR is switched ON in addition to MPI.

\subsection{\label{subsec:DepFg}Effect of change in gluonic contribution on observed jet modification}
\begin{figure}[!h]
	\centering
	\includegraphics[scale=.45]{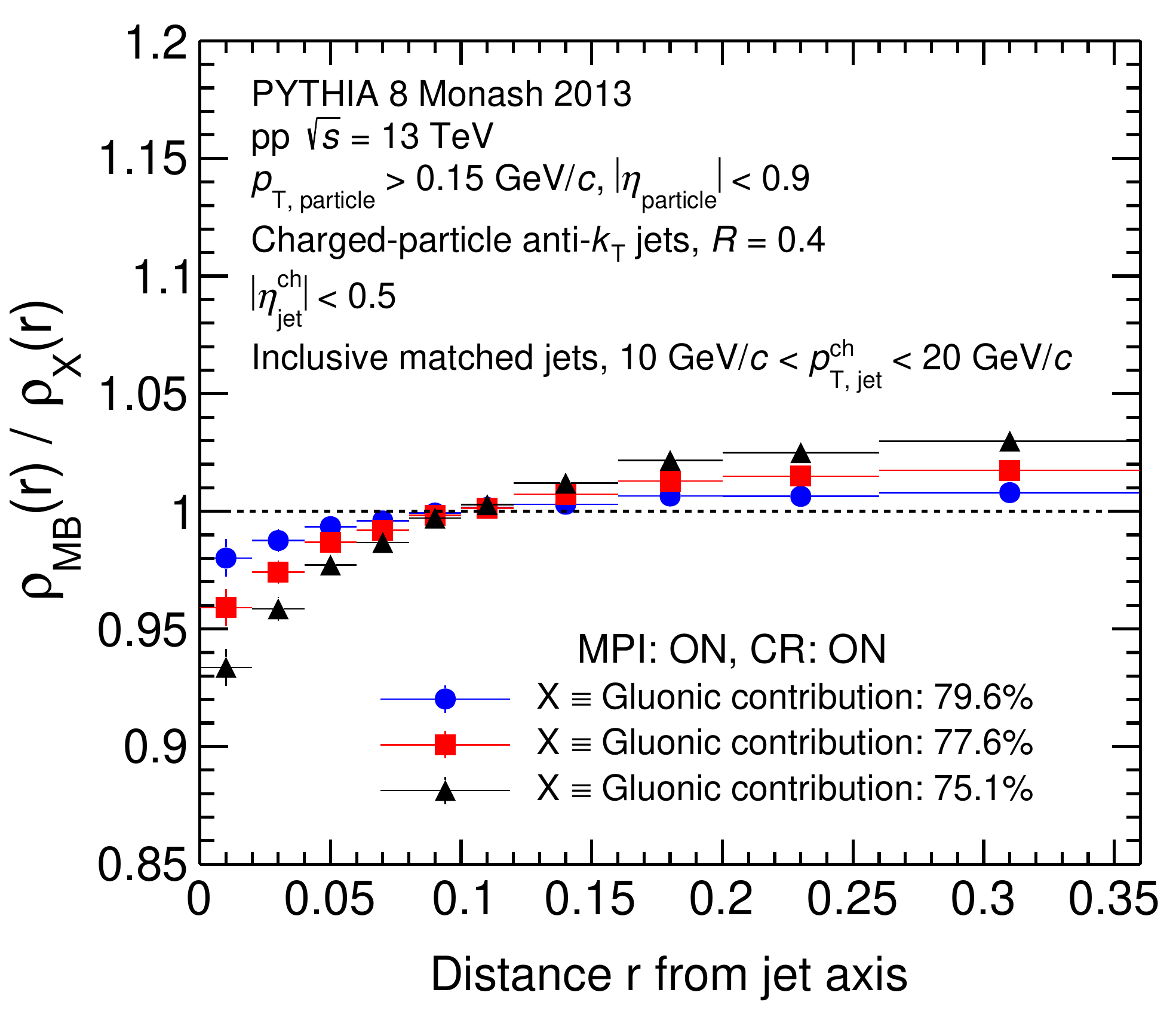}
	\caption{Ratio of $\rho(r)$ distributions obtained from sample-II, sample-III and sample-IV to that in sample-I (MB) in `MPI:~ON, CR:~ON' configuration.}
	\label{refFg}
\end{figure}
In order to understand exclusively the effect of change in gluonic contribution on jet modification, we study $\rho(r)$ distribution in the interval 10~$<p_{\rm T,\,jet}^{\rm ch}<$~20~GeV/$c$ for jet samples containing different fractions of gluon-initiated jets. We perform this study using PYTHIA~8 with default configuration (`MPI:~ON, CR:~ON'). We first consider inclusive matched jets from MB events as sample-I (where the gluonic contribution is 81.2\%), then we randomly remove 10\%, 20\% and 30\% of gluon-initiated jets from sample-I to obtain sample-II, sample-III and sample-IV respectively which correspond to 79.6\%, 77.6\% and 75.1\% gluonic contributions in inclusive matched jets. To illustrate the sensitivity of jet modification to the change in the fraction of gluon-initiated jets, we study the ratio of $\rho(r)$ distributions obtained from sample-II, III and IV to that from sample-I as shown in Fig.~\ref{refFg}. As expected, the modification in $\rho(r)$ increases with increasing change in gluonic contribution. It is found that a change of 6\% in the gluonic contribution from 81.2\% (sample-I) to 75.1\% (sample-IV) leads to about 7\% modification of jet core ($\left.\rho(r)\right|_{r \rightarrow 0}$). This observation further supports our finding (as discussed in Sec.~\ref{sec:ResJetShapeAndFragmentation}) that the amount of jet modification indeed is dependent on the change in gluonic contribution.

\section{\label{sec:Summary}Summary}
In summary, we studied the differential jet transverse momentum, differential jet shape $\rho(r)$ and jet fragmentation $z^{\rm ch}$ distributions for inclusive charged-particle jets in high-multiplicity and minimum bias pp collisions at $\sqrt{s}$~=~13~TeV using PYTHIA~8 Monash~2013 MC simulation. Jets were reconstructed from charged particles at mid-rapidity using the anti-$k_{\rm T}$ jet finding algorithm with jet resolution parameter $R$~=~0.4. Contributions from UE were estimated using the perpendicular cone method and subtracted on a statistical basis. The distributions of $\rho(r)$ and $z^{\rm ch}$ were obtained with `MPI:~OFF, CR:~OFF', `MPI:~ON, CR:~OFF' and `MPI:~ON, CR:~ON' configurations. HM event class was defined as the top 5\% of the total events with the highest multiplicities, based on the number of charged particles produced in the pseudorapidity range 2.8~$\textless~\eta~\textless$~5.1  and $-\rm{3.7}~\textless~\eta~\textless~-$1.7.
\par
We observed a significant amount of modification of jet shape $\rho(r)$ and jet fragmentation $z^{\rm ch}$ distributions in HM event class compared to MB event class for 10~$<p_{\rm T,\,jet}^{\rm ch}<$~20~GeV/$c$. The jet core ($\left.\rho(r)\right|_{r \rightarrow 0}$) is modified by about 22\% followed by a small enhancement at larger $r$~($>0.15$) whereas the production of high $z^{\rm ch}$ particles is suppressed by about 40\% at $z^{\rm ch}\rightarrow$~1 in HM event class compared to that in MB. The enhanced number of MPIs and the change in the number of gluon-initiated jets in HM event class along with the color reconnection mechanism are the main sources causing modifications of $\rho(r)$ and $z^{\rm ch}$ distributions. For high-$p_{\rm T}$ jets (40~$<p_{\rm T,\,jet}^{\rm ch}<$~60~GeV/$c$), the observed jet modification is significantly reduced.
We also find a direct connection of $\langle N_{\rm MPI}\rangle$ and gluonic contribution with the amount of modification in $\rho(r)$ -- the larger the number of MPIs and/or gluonic contribution, the larger the amount of modification of $\rho(r)$. 
\par
The findings of this work emphasize the necessity of a better understanding of the origin of particle production in high-multiplicity events in pp collisions. The modification of $\rho(r)$ and $z^{\rm ch}$ in HM events and their strong correlations with the underlying physics mechanisms such as MPI and CR as well as with the change in the gluonic contribution compared to that in MB events demand a very careful study and interpretation of such observables by the experimental community. Although disentangling gluon-initiated jets experimentally will be challenging, however, studying the multiplicity dependence of $\rho(r)$ and $z^{\rm ch}$ for pure gluonic jets would be worth pursuing.

\section*{Acknowledgements}
The authors thank Prof. Claude Pruneau for useful discussions.
D. Banerjee acknowledges the Inspire Fellowship research grant [DST/INSPIRE Fellowship/2018/IF180285]. Significant part of computation for this work was carried out using the computing server facility at CAPSS, Bose Institute, Kolkata.

\section*{References}
\bibliographystyle{utphys} 
\bibliography{JetModificationInppPythia8}

\end{document}